\documentclass[journal]{IEEEtran}

\usepackage{booktabs}
\usepackage{tabularx}

\usepackage{amssymb}
\usepackage{amsbsy}
\usepackage{amsmath}
\usepackage{bm}
\usepackage{verbatim}
\usepackage{mathrsfs}
\usepackage{amsfonts}
\usepackage{graphicx}
\usepackage[tight,footnotesize]{subfigure}
\usepackage[10pt]{moresize}
\usepackage{array}
\usepackage{color}
\usepackage{epsfig}
\usepackage{stfloats}
\usepackage{balance}

\usepackage[colorlinks]{hyperref}

\usepackage[noend]{algpseudocode}
\usepackage{algorithmicx,algorithm}

\usepackage{setspace}
\usepackage{cases}

\usepackage{epstopdf}
\usepackage{multirow}
\usepackage{extarrows}

\usepackage{pifont} 
\usepackage{enumerate}
\usepackage{enumitem}

\newcommand{\subparagraph}{}
\usepackage{titlesec}
\titlespacing{\section}{0pt}{2 ex plus .0ex minus .0ex}{1ex plus .0ex}
\titlespacing{\subsection}{0pt}{1.5 ex plus .0ex minus .0ex}{0.8 ex plus 0.0ex}
\titlespacing{\subsubsection}{0pt}{0.5ex plus .0ex minus .0ex}{0.0ex plus .0ex}

\allowdisplaybreaks[4]

\ifCLASSINFOpdf

\else

\fi

\begin{document}

\title{Orchestrating Multimodal DNN Workloads in Wireless Neural Processing}

\author{Sai Xu,~\IEEEmembership{Member,~IEEE}, 
        Kai-Kit Wong,~\IEEEmembership{Fellow,~IEEE}, 
        Yanan~Du,~\IEEEmembership{Member,~IEEE},
        Hyundong Shin,~\IEEEmembership{Fellow,~IEEE}
\thanks{S. Xu and K. K. Wong are with the Department of Electronic and Electrical Engineering, University College London, WC1E 7JE, London, UK. K. K. Wong is also with the Department of Electronic Engineering, Kyung Hee University, Yongin-si, Gyeonggi-do 17104, Republic of Korea. (e-mail: $\rm\{sai.xu,kai\text{-}kit.wong\}@ucl.ac.uk$). Y. Du is with the School of Electrical and Electronic Engineering, the University of Sheffield, Sheffield S10 2TN, UK. (e-mail: $\rm yanan.du@sheffield.ac.uk$). H. Shin is with the Department of Electronics and Information Convergence Engineering, Kyung Hee University, Yongin-si, Gyeonggi-do 17104, Republic of Korea (e-mail: $\rm hshin@khu.ac.kr$).}
\thanks{Corresponding author: Kai-Kit Wong.}
}

\maketitle

\begin{abstract}
In edge inference, wireless resource allocation and accelerator-level deep neural network (DNN) scheduling have yet to be co-optimized in an end-to-end manner. The lack of coordination between wireless transmission and accelerator-level DNN execution prevents efficient overlap, leading to higher end-to-end inference latency. To address this issue, this paper investigates multimodal DNN workload orchestration in wireless neural processing (WNP), a paradigm that integrates wireless transmission and multi-core accelerator execution into a unified end-to-end pipeline. First, we develop a unified communication–computation model for multimodal DNN execution and formulate the corresponding optimization problem. Second, we propose O-WiN, a framework that orchestrates DNN workloads in WNP through two tightly coupled stages: simulation-based optimization and runtime execution. Third, we develop two algorithms, RTFS and PACS. RTFS schedules communication and computation sequentially, whereas PACS interleaves them to enable pipeline parallelism by overlapping wireless data transfer with accelerator-level DNN execution. Simulation results demonstrate that PACS significantly outperforms RTFS under high modality heterogeneity by better masking wireless latency through communication–computation overlap, thereby highlighting the effectiveness of communication–computation pipelining in accelerating multimodal DNN execution in WNP.
\end{abstract}


\IEEEpeerreviewmaketitle

\section{Introduction}\label{sec:introduction}
\IEEEPARstart{I}{MAGINE} a scenario where diverse data streams from multiple distributed sensor nodes are wirelessly offloaded to an edge server. Once collected, these heterogeneous streams are fed into a single multimodal deep neural network (DNN) for inference~\cite{Xu2022Multimodal}. Because the end-to-end inference latency is jointly determined by the DNN execution time on the computing platform and the wireless transmission time, communication is as critical as computation~\cite{Wen2024Task,Yi2025Towards}; moreover, under limited spectrum resources, the overall latency is frequently wireless-bound in some scenarios (as shown in Fig.~\ref{Fig6}). However, wireless resource allocation and accelerator-level DNN scheduling are typically designed in isolation. More precisely, they are coupled and coordinated only via an overlay framework, rather than through underlying end-to-end cross-domain co-optimization~\cite{Wen2024Task, Kundu2026AI-RAN}. Motivated by this observation, this paper studies wireless neural processing (WNP), a new paradigm that integrates wireless transmission with DNN execution on multi-core accelerators to enable fine-grained pipeline parallelism across communication and computation.

WNP seeks to unify wireless communication and neural processing into a single execution pipeline, moving beyond the traditional view of wireless communication as merely an external I/O operation. In this unified execution flow, data transfer is tightly interleaved with neural processing at the operator level. This temporal parallelism distinguishes WNP from existing communication–computation co-design approaches~\cite{Shi2020Communication}, which rely on coarse-grained resource collaboration. As a result, WNP could help alleviate the “wireless wall” and deliver lower end-to-end latency, higher throughput, and improved energy efficiency for DNN workloads. As a new research area, a fundamental question in WNP is how to orchestrate DNN workloads such that wireless data transmission and multi-core accelerator execution are temporally aligned, thereby enabling seamless end-to-end communication–computation integration.

\begin{table*}[t]
\scriptsize
\caption{Main Notation}
\label{tab:notation}
\centering
\setlength{\tabcolsep}{4pt}
\renewcommand{\arraystretch}{1.05}
\begin{tabularx}{\textwidth}{@{}lXlX@{}}
\toprule
\textbf{Symbol} & \textbf{Description} & \textbf{Symbol} & \textbf{Description} \\
\midrule
$\mathcal{G}=(\mathcal{V},\mathcal{E})$ & Computation DAG of the multimodal DNN & $\mathcal{V},\ \mathcal{E}$ & Vertex set and edge set of $\mathcal{G}$ \\
$\mathrm{pred}(v),\ \mathrm{succ}(v)$ & Predecessor/successor sets of job $v$ & $(u\!\to\!v)\in\mathcal{E}$ & Precedence constraint: $v$ can start only after $u$ finishes \\
$\mathcal{K}=\{1,\ldots,K\}$ & Set of modalities (sensor nodes) & $\mathcal{G}_k=(\mathcal{V}_k,\mathcal{E}_k)$ & Subgraph for modality $k$ \\
$\mathcal{C}=\{1,\ldots,C\}$ & Set of accelerator cores & $s_v,\ f_v$ & Start/finish time of job $v$ \\
$y_{v,c}\in\{0,1\}$ & Job-to-core assignment indicator & $p_{v,c}$ & Processing time of job $v$ on core $c$ \\
$\mathcal{F}_c$ & Feasible schedules on core $c$ & $B_{\mathrm{tot}}$ & Total NoC bandwidth budget \\
$b_c(t)$ & Bandwidth usage of core $c$ at time $t$ & $\mathcal{T}=\{0,\ldots,T-1\}$ & Set of wireless time slots \\
$\mathcal{F}=\{1,\ldots,F\}$ & Set of OFDMA subcarriers & $\Delta$ & Time-slot duration \\
$x_{k,f,t}\in\{0,1\}$ & RB assignment: node $k$ uses RB $f$ at slot $t$ & $\rho_{k,f,t}$ & Uplink rate of node $k$ on RB $f$ at slot $t$ \\
$D_k$ & Slice size of modality $k$ & $B_k(t;\mathbf{x})$ & Cumulative received bits of node $k$ by end of slot $t$ \\
$t_k^{\mathrm{arr}}(\mathbf{x})$ & Arrival slot index of slice $k$ at the edge server & $r_k(\mathbf{x})$ & Arrival (release) time of slice $k$ at the edge server \\
$T_{\mathrm{makespan}}$ & End-to-end inference makespan & $t_{\mathrm{start}}$ & Compute start time in RTFS (wait-all) \\
$R_k(t;\mathbf{x})$ & Remaining payload of slice $k$ at the beginning of slot $t$ & $\hat{u}_k(t)$ & Predicted effective uplink rate of node $k$ at slot $t$ \\
$\bar{u}_k(t),\ u_{k,f}(t)$ & Mean/instantaneous rate of node $k$ at slot $t$ & $\hat{T}_k(t)$ & Predicted residual transmission time of slice $k$ (RTFS) \\
$L_{v,c},\ b_{v,c}$ & Profiled latency and bandwidth demand of job $v$ on core $c$ & $B_{\max}$ & Per-core NoC bandwidth cap \\
$\overline{L}_v$ & Average latency of job $v$ across cores & $\mathrm{rank}_u(v)$ & Upward rank in list scheduling \\
$\sigma(v)$ & Job state in the simulator (unscheduled/running/finished) & $\mathcal{Q}_{\mathrm{ready}}$ & Ready-job queue \\
$v_c,\ \omega_c$ & Running job on core $c$ and its remaining work & $\alpha_c$ & Bandwidth share allocated to core $c$ (event step) \\
$\delta$ & Event-step duration until the next job completion & $\mathcal{P}$ & Lightweight predictor (PACS) \\
$\hat{A}_k(t)$ & Predicted arrival slot of slice $k$ at time $t$ (PACS) & $S_v(t)$ & Predictor release time for subgraph $\mathcal{V}_k$ (PACS) \\
$\hat{F}_v(t)$ & Predicted earliest finish time of job $v$ (PACS) & $o$ & Output node of the DAG \\
$\epsilon$ & \multicolumn{2}{l}{Small positive constant (numerical stability)} & \\
$G(k)$ & \multicolumn{2}{l}{Predicted makespan reduction by assigning one RB to slice $k$ (PACS)} &  \\
\bottomrule
\end{tabularx}
\end{table*}

\subsection{Related Work}\label{sec:relatedwork}

This work is inherently interdisciplinary, which bridges edge artificial intelligence (AI) inference and accelerator-level DNN execution. Accordingly, we organize the related literature into two categories and review the state of the art in each.

\textit{1) Edge AI Inference:} From the perspective of edge AI inference, DNN inference is often treated under a compute-as-a-service (CaaS) abstraction: the DNN is modeled as a schedulable workload, while accelerator-level operator execution and system constraints (e.g., memory and bandwidth) receive limited attention. In edge AI, collaborative inference has become a prominent research direction. The key idea is to partition large DNN models in a layer-wise or graph-wise manner, and to execute the resulting segments jointly on end devices and computing centers~\cite{Zeng2020Edge}. To reduce uplink communication overhead, intermediate activations or feature representations at the partition point are typically transmitted instead of raw inputs~\cite{Yang2020Energy,Li2023Throughput}. Since uplink communication cost typically scales with the size of the transmitted feature representations, integrating feature compression can substantially reduce communication overhead and enable low-latency edge inference~\cite{Shao2021Learning,Shao2020Bottlenet,Xiao2024Adaptive,FrankenSplit2024Furutanpey}. To further mitigate the communication bottleneck in device–edge collaborative inference, task-oriented communication~\cite{Shao2022Task,Xie2023Robust,Li2025Tackling,Xie2025Toward} has been developed to move beyond faithful data delivery and instead transmit only the information necessary for task completion, thereby reducing uplink overhead and latency while maintaining task performance and improving robustness.

\textit{2) Accelerator-Level DNN Execution:} In the computer architecture community, an active research direction in recent years is efficient execution of heterogeneous DNN workloads on multi-core accelerators. Generally, the scheduling design space is enormous, spanning both operator execution order and core-assignment decisions. To obtain high-quality schedules with low computational complexity, several scheduling strategies have been proposed. In data-center settings, AI-MT~\cite{AI-MT} adopted a hand-crafted layer-level scheduler for multi-DNN inference. MAGMA~\cite{Magma} and COMB~\cite{COMB} employed genetic algorithms to speed up schedule exploration, with COMB further incorporating memory-aware optimizations. MoCA~\cite{MoCA} improved multi-DNN performance through dynamic memory management. For edge scenarios, Herald~\cite{Herald} and DREAM~\cite{DREAM} targeted real-time multi-DNN execution; Herald leveraged a heterogeneous dataflow design, whereas DREAM used an adaptive online scheduler to cope with time-varying workloads. DySta~\cite{DySta} was tailored for sparse multi-DNN workloads. TaiChi~\cite{TaiChi} proposed a framework that combines graph neural networks and reinforcement learning for efficient multi-DNN scheduling on a multi-core accelerator.

Prior work has either decoupled edge AI inference from DNN execution, or limited communication–computation co-design to coarse-grained, high-level coordination. From a system perspective, the wireless channel fundamentally serves as the data-supply path during inference. Functionally, it can be abstracted as a bandwidth-limited, high-latency, and dynamically varying form of “remote memory” for a deep-learning accelerator. Therefore, we argue that wireless data streams should be incorporated into a unified dataflow modeling and optimization framework for accelerator-level DNN execution—an approach that is highly promising for further improving edge AI inference performance.

\subsection{Contributions}
To bridge the research gap in jointly optimizing wireless resource allocation and accelerator-level DNN execution and enable systematic exploration of WNP, this paper explores a comprehensive solution that orchestrates multimodal DNN workloads through a unified communication–computation pipeline. Our main contributions are summarized as follows:
\begin{itemize}
\item We build a system architecture for multimodal DNN execution in WNP that enables joint consideration of wireless data delivery and operator-level execution on multi-core accelerators. This architecture captures the underlying interactions between wireless transmission and operator-level DNN execution. Building on this model, we formulate an optimization problem that aims to minimize the total end-to-end inference latency for multimodal communication–computation orchestration in WNP.
\item We present O-WiN, a modular and scalable orchestration framework that decomposes the unified communication–computation pipeline into interface-defined functional modules for decoupled integration and systematic cross-module coordination. O-WiN features two tightly coupled stages—simulation-based optimization and runtime execution—where the simulation stage iteratively refines communication policies and computation mapping strategies, and the resulting solutions are deployed to optimize wireless transmission and accelerator execution at runtime.
\item To tackle the end-to-end inference latency minimization problem in WNP, we develop two algorithms: (i) RTFS, a sequential communication–computation heuristic that prioritizes uplink delivery to minimize wireless transmission latency before inference; and (ii) PACS, a pipelined heuristic that overlaps wireless delivery with operator execution, triggering the inference of a modality-specific branch as soon as the minimum required input data for that modality becomes available.
\item We conduct simulations to evaluate RTFS and PACS, reporting end-to-end execution timelines and per-core network-on-chip (NoC)  bandwidth allocation. We further perform sensitivity studies by varying the number of accelerator cores, orthogonal frequency-division multiple access (OFDMA) subcarriers, wireless latency scaling factor, NoC bandwidth budget, as well as compression factors and slice token sizes.
\end{itemize}

The rest of the paper is organized as follows. Section~\ref{sec:model} presents the system model and formulates the multimodal communication--computation orchestration problem in WNP. Section~\ref{sec:framework} proposes the framework O-WiN and details the functionality of each module and the overall workflow. Section~\ref{sec:algorithms} introduces two heuristic algorithms to enable stage-wise execution and pipeline parallelism, respectively. Section~\ref{sec:evaluation} reports simulation results and discusses key performance trends under different system configurations. Finally, Section~\ref{sec:conclusions} concludes the paper.

\section{System Model and Problem Formulation}\label{sec:model}
\begin{figure*}
\centering
\includegraphics[width=\linewidth]{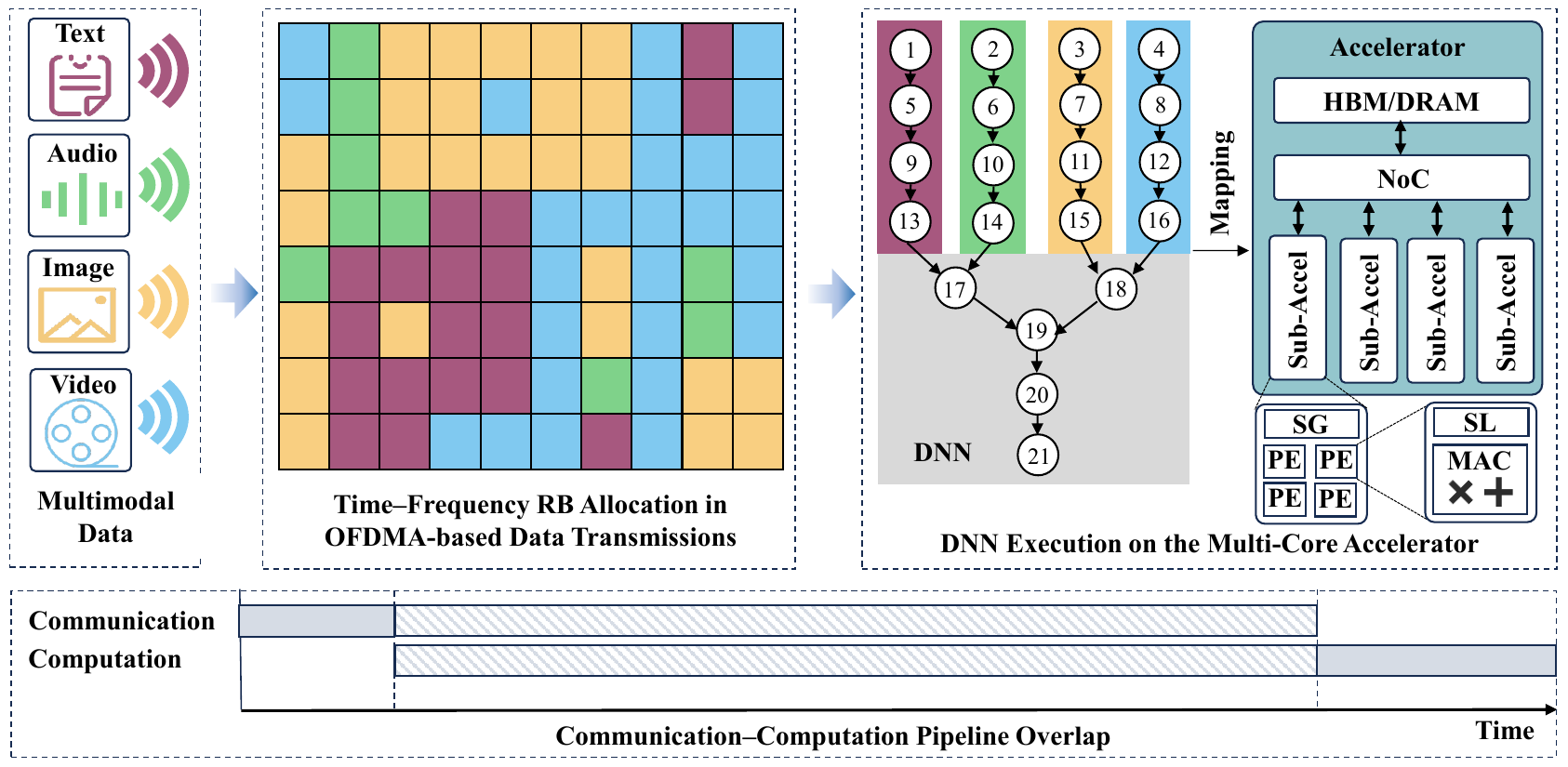}
\caption{A system architecture for multimodal DNN execution in WNP.}\label{Fig1}
\end{figure*}
Fig. \ref{Fig1} illustrates a system architecture for multimodal DNN execution in WNP. In the system, the multimodal DNN inputs are transmitted from sensor nodes to an edge server over wireless uplink channels. Upon data arrival at the edge server, the inference of multimodal DNN model is triggered, and its operators are mapped onto a multi-core accelerator for execution. To accelerate the inference, pipeline parallelism across communication and computation can be leveraged to hide wireless transmission latency. Table \ref{tab:notation} summarizes the adopted main notations in this paper.

\subsection{Neural Network Architecture}
At the edge server, a trained multimodal DNN is deployed to process heterogeneous data collected from multiple sensor nodes, including text, audio, images, and video. The model typically consists of modality-specific encoders that map each modality input to a latent feature representation. These features are then integrated via a fusion mechanism, such as concatenation followed by projection, attention-based fusion, or gating, to obtain a unified representation. The unified representation is subsequently fed into task-specific heads to produce predictions. When different modalities provide complementary and well-aligned information, multimodal DNNs often achieve higher accuracy and robustness than single-modality models.\par

The multimodal DNN inference workload is modeled as a directed acyclic graph (DAG) denoted by $\mathcal{G}=(\mathcal{V},\mathcal{E})$, where each vertex $v\in\mathcal{V}$ represents a computational operator and each directed edge $(u\!\rightarrow\! v)\in\mathcal{E}$ captures a data dependency, meaning that operator $v$ can start only after receiving the output of operator $u$. The operators are grouped into modality-specific encoder subgraphs, each processing inputs from one modality, while additional subgraphs may represent fusion and task heads. Let $\mathcal{K}=\{1,\ldots,K\}$ and $\mathcal{G}_k=(\mathcal{V}_k,\mathcal{E}_k)$ denote the index set of modalities and the modality-specific subgraph corresponding to modality $k\in\mathcal{K}$, respectively.

\subsection{Multi-Core Accelerator}
The accelerator consists of multiple cores, referred to as sub-accelerators, arranged in a processing element (PE) array. All cores access the shared system bandwidth through the NoC, where the system bandwidth is defined as the minimum of the main-memory (e.g., HBM/DRAM) bandwidth and the host-to-accelerator (e.g., PCIe) bandwidth. Each PE contains a multiply–accumulate (MAC) unit for computing partial sums and a small local scratchpad (SL) for storing weights, activations, and intermediate results. At the core level, a shared global cache (SG) prefetches data required for the next computation tile from HBM/DRAM. The NoC distributes these prefetched operands to the PEs’ SLs, and computation results are written back from SL to SG. \par

To model how layer-level computations are executed on a multi-core accelerator, the notions of job and mapping are defined as follows: (i) A job is the smallest computation task that can be independently scheduled and executed on the underlying hardware accelerator. A job may correspond to an entire layer or a partitioned portion of a layer, depending on the layer's size and the available hardware resources. It serves as the fundamental unit for scheduling and computation in the system. (ii) A mapping is the strategy for orchestrating the assignment, scheduling, and prioritization of jobs across multiple cores in both space and time. It defines how computation and data movement are coordinated on the hardware to achieve efficient resource utilization and optimized performance. Mapping strategies can be manually designed or generated automatically through optimization algorithms.\par

Let $\mathcal{C}=\{1,\dots,C\}$ denote the set of accelerator cores. For each job $v\in\mathcal{V}$, let $s_v$ and $f_v$ denote its start and finish times, respectively. We introduce binary assignment variables $y_{v,c}\in\{0,1\}$, where $y_{v,c}=1$ indicates that job $v$ is assigned to core $c\in\mathcal{C}$. Each job must be assigned to exactly one core:
\begin{equation}
\sum_{c\in\mathcal{C}} y_{v,c}=1,\qquad \forall v\in\mathcal{V}.
\label{eq:core_assignment}
\end{equation}
If executed on core $c$, job $v$ requires a processing time of $p_{v,c}$, and jobs are assumed to be non-preemptive. Accordingly, the finish time of job $v$ is
\begin{equation}
f_v = s_v + \sum_{c\in\mathcal{C}} y_{v,c}\,p_{v,c},\quad \forall v\in\mathcal{V}.
\label{eq:finish_time}
\end{equation}
Due to DAG precedence constraints, a job can start only after all its predecessors finish:
\begin{equation}
s_v \ge f_u,\qquad \forall (u,v)\in \mathcal{E}.
\label{eq:dag_precedence}
\end{equation}
Because of hardware limitations, a core can execute at most one job at a time. To avoid introducing additional integer variables and big-$M$ linearization, we impose the core-capacity constraint in a feasibility form:
\begin{equation}
\begin{aligned}
\{(s_v,f_v)\mid y_{v,c}=1,\ v\in\mathcal{V}\}\ \in \mathcal{F}_c, \forall c\in\mathcal{C},
\end{aligned}
\label{eq:nonoverlap}
\end{equation}
where $\mathcal{F}_c$ denotes the set of feasible schedules on core $c$ that satisfy the disjunctive non-overlap constraints. Equivalently, for any two distinct jobs $u\neq v$ with $y_{u,c}=y_{v,c}=1$, either $f_u\le s_v$ or $f_v\le s_u$ must hold.

In a multi-core accelerator, NoC bandwidth is shared across cores and can be dynamically redistributed. The total bandwidth usage of all cores at any time must not exceed the available budget:
\begin{equation}
\sum_{c\in\mathcal{C}} b_c(t)\le B_{\mathrm{tot}},\quad \forall t.
\label{eq:bandwidth_global}
\end{equation}
where $B_{\mathrm{tot}}$ and $b_c(t)\ge 0$ denote the total bandwidth and the bandwidth usage of core $c\in\mathcal{C}$ at time $t$, respectively.

\subsection{Wireless Transmission}
To support simultaneous and efficient multi-node access, OFDMA is adopted. In OFDMA, the available spectrum is partitioned into a set of orthogonal subcarriers and tiled over the time axis, forming a grid of time–frequency resource blocks (RBs). Due to location-dependent channel gains and spatially heterogeneous interference, different nodes experience different uplink transmission rates on the same RB; moreover, the rate of a given node can vary significantly across different RBs. To maximize spectral efficiency under time-varying channel conditions, RBs are dynamically allocated among nodes.

Let $\mathcal{T}=\{0,\ldots,T-1\}$ and $\mathcal{F}=\{1,\ldots, F\}$ denote the sets of time slots and orthogonal subcarriers, respectively. Define a slice as the smallest unit of data generated from a single modality that is sufficient to initiate the forward inference of its corresponding modality-specific subgraph. As the fundamental processing unit, a slice can contain either a portion of the modality data or the entire data instance, where its size is determined by the input interface of the encoder. Assume that each sensor node is associated with a unique modality and use $k\in\mathcal{K}$ to index the sensor node corresponding to modality $k$, yielding a one-to-one mapping between nodes and modalities. A binary variable $x_{k,f,t}\in\{0,1\}$ indicates the RB allocation decision: $x_{k,f,t}=1$ if RB $f$ is assigned to node $k$ in slot $t$, and $0$ otherwise. Since each RB can be assigned to at most one modality for uplink transmission, the following exclusivity constraint must hold:
\begin{equation}
\sum_{k\in\mathcal{K}} x_{k,f,t}\le 1,\qquad \forall f\in\mathcal{F},\ \forall t\in\mathcal{T}.
\label{eq:rb_exclusivity_slot}
\end{equation}
Let $\Delta$ and $\rho_{k,f,t}$ denote the slot duration and the achievable uplink rate of node $k$ on RB $f$ in slot $t$, respectively. Node $k$ can send $\Delta\,\rho_{k,f,t}$ bits on RB $f$ during slot $t$. Given an RB allocation $\mathbf{x}=\{x_{k,f,t}\}$, the total number of bits received from node $k$ at the edge server by the end of slot $t$ is
\begin{equation}
B_k(t;\mathbf{x})=\sum_{\tau=1}^{t}\ \sum_{f\in\mathcal{F}}
x_{k,f,\tau}\,\Delta\,\rho_{k,f,\tau}.
\label{eq:cum_bits_k}
\end{equation}
Let $D_k$ denote the slice size of node $k$. The arrival slot of the slice is given by
\begin{equation}
t_k^{\mathrm{arr}}(\mathbf{x})=\min\Bigl\{t\in\mathcal{T}:\ B_k(t;\mathbf{x})\ge D_k\Bigr\}.
\label{eq:arr_slot_k}
\end{equation}
The corresponding arrival time at the edge server is given by
\begin{equation}
r_k(\mathbf{x}) = t_k^{\mathrm{arr}}(\mathbf{x})\,\Delta.
\label{eq:arrival_time_slot_k}
\end{equation}
As the arrival time at the edge server also serves as the release time of its associated modality-specific subgraph, no operator in $\mathcal{G}_k=(\mathcal{V}_k,\mathcal{E}_k)$ can start before $r_k(\mathbf{x})$. 
\begin{equation}
s_v \ge r_k(\mathbf{x}),\qquad \forall k\in\mathcal{K},\ \forall v \in \mathcal{V}_k.
\label{eq:cross_plane_release}
\end{equation}
\subsection{Parallelism}
A complete multimodal inference process depends not only on how DNN layers are mapped onto the hardware accelerator, but also on the timing of input data arrivals. As a result, the key coupling between wireless transmission and DNN execution lies in data arrival time. In a multimodal DNN, each modality-specific subgraph relies solely on its corresponding input data and can start execution immediately once the data become available. In other words, once a modality arrives, its corresponding subgraph can execute even while other modalities are still being transmitted. For instance, when a subgraph has a large computational workload but a relatively small input data size, the execution is computation-bound, and prioritizing the transmission of its modality data may effectively reduce overall latency. Conversely, when a subgraph involves a small computational workload but a large input data size, deferring its data transmission may be more efficient. Taking these factors into account, enhancing the parallelism of the communication–computation pipeline and avoiding accelerator underutilization require well-coordinated communication–computation resource allocation and task scheduling.
\subsection{Problem Formulation}\label{sec:formulation}
During the inference process, communication and computation form a unified execution pipeline. We adopt the end-to-end inference latency (i.e., the makespan) as the performance metric, given by
\begin{equation}
T_{\mathrm{makespan}} = \max_{v \in \mathcal{V}}~f_v .
\end{equation}
Our goal is to minimize the makespan by jointly optimizing communication-side resource allocation, computation-side resource allocation, and job scheduling. The optimization problem can be formulated as
\begin{align}
\min_{\mathbf{x},\mathbf{y},\mathbf{s},\mathbf{f}}\quad & T_{\mathrm{makespan}} \label{eq:obj}\\
\text{s.t.}\quad
& \eqref{eq:core_assignment},\ \eqref{eq:finish_time},\ \eqref{eq:dag_precedence},\ \eqref{eq:nonoverlap},\ \eqref{eq:bandwidth_global},\ \eqref{eq:rb_exclusivity_slot},\ \eqref{eq:cross_plane_release},
\end{align}
where $\mathbf{y}=\{y_{v,c}\}_{v\in\mathcal{V},\,c\in\mathcal{C}}$, $\mathbf{s}=\{s_v\}_{v\in\mathcal{V}}$ and $\mathbf{f}=\{f_v\}_{v\in\mathcal{V}}$.

\section{Framework}\label{sec:framework}
\begin{figure}
\centering
\includegraphics[width=\linewidth]{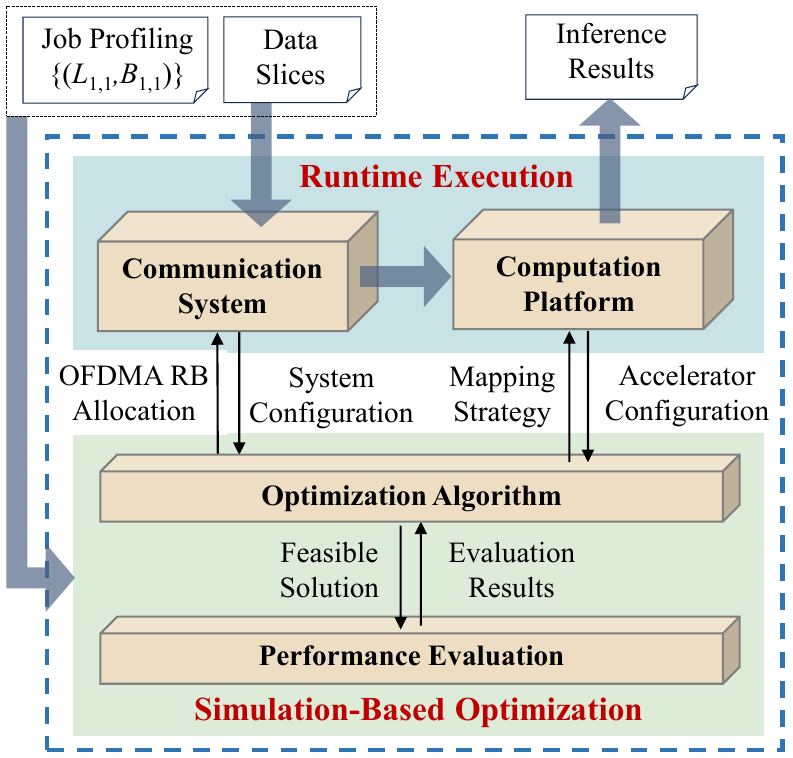}
\caption{O-WiN: A modular and scalable orchestration framework that decomposes the unified communication–computation pipeline into interface-defined functional modules for decoupled integration and systematic cross-module coordination.}\label{Fig2} 
\end{figure}

To enable multimodal DNN workload orchestration in WNP, it is essential to establish a unified and reusable abstraction that spans the entire pipeline---from data preparation and algorithm development to system execution and performance evaluation. In this section, we introduce O-WiN, a modular and scalable orchestration framework that decomposes the end-to-end workflow into well-defined, interface-driven functional modules, enabling decoupled integration, systematic coordination, and future extensibility of each component. The overall architecture of O-WiN is illustrated in Fig.~\ref{Fig2}.\par

\subsection{Architecture}\label{sec:communicationmodule}
O-WiN consists of four decoupled, plug-and-play modules with well-specified interfaces:
\begin{itemize}
    \item \textit{Communication System.} Given the RB allocation decision $\{x_{k,f,t}\}$, the per-RB uplink rate $\{\rho_{k,f,t}\}$, and the slice size $\{D_k\}$, this module determines when each slice is completely received at the edge server and outputs its arrival (release) time $\{r_k\}$. Moreover, it delivers the received slice payload to the downstream computation platform module for subsequent inference.
    \item \textit{Computation Platform.} Given the received slices and their release times $\{r_k\}$, the multimodal inference DAG $\mathcal{G}=(\mathcal{V},\mathcal{E})$, and a mapping strategy comprising the job-to-core assignment $\{y_{v,c}\}$ and a per-core execution schedule $\{\pi_c\}$ with $\pi_c\in\mathcal{F}_c$, this module executes inference, produces the outputs at the sink node $o$, and reports the resulting end-to-end makespan $T_{\mathrm{makespan}}$.
    \item \textit{Optimization Algorithm.} Given the slice size $\{D_k\}$, the per-RB uplink rate $\{\rho_{k,f,t}\}$, the multimodal inference DAG $\mathcal{G}=(\mathcal{V},\mathcal{E})$, the job profiling table $\{(L_{v,c},\, b_{v,c})\}$\footnote{The job profiling table $\{(L_{v,c},\, b_{v,c})\}$ can be constructed from (i) empirical profiling on real hardware, (ii) measurements obtained via a cycle-/trace-level simulator, or (iii) synthetically generated workloads (e.g., random sampling from pre-defined distributions).}, and the system configurations on both communication and computation, this module performs communication--computation co-optimization to minimize the end-to-end makespan $T_{\mathrm{makespan}}$. It returns (i) an RB allocation decision $\{x_{k,f,t}\}$ and (ii) a mapping policy, comprising the job-to-core assignment $\{y_{v,c}\}$ and a per-core execution schedule $\{\pi_c\}$ with $\pi_c\in\mathcal{F}_c$.
    \item \textit{Performance Evaluation.} Given the slice size $\{D_k\}$, the per-RB uplink rate $\{\rho_{k,f,t}\}$, the multimodal inference DAG $\mathcal{G}=(\mathcal{V},\mathcal{E})$, the job profiling table $\{(L_{v,c},\, b_{v,c})\}$, the system configurations on both communication and computation, and a feasible co-optimization solution $\big(\{x_{k,f,t}\},\,\{y_{v,c}\},\,\{\pi_c\}\big)$, the module executes a full-stack simulation that emulates slice transmissions to obtain $\{r_k\}$, runs DAG execution under the induced release times and schedules, and reports end-to-end performance metrics, including $T_{\mathrm{makespan}}$ and other relevant statistics (e.g., core utilization and NoC bandwidth usage). The ultimate output is the final solution $\big(\{x^{*}_{k,f,t}\},\,\{y^{*}_{v,c}\},\,\{\pi^{*}_c\}\big)$ and its validated evaluation results.
\end{itemize}

\subsection{Workflow}
The end-to-end workflow of O-WiN comprises two tightly coupled stages: simulation-based optimization and runtime execution. The simulation stage efficiently searches for a high-quality co-optimization solution $\big(\{x^{*}_{k,f,t}\},\,\{y^{*}_{v,c}\},\,\{\pi^{*}_c\}\big)$, which is then deployed to drive the runtime inference pipeline.

\textit{1) Simulation-Based Optimization:} O-WiN adopts an iterative optimize--evaluate loop that tightly couples the optimization-algorithm module with the performance-evaluation module. In each iteration, the optimizer generates a candidate co-optimization policy, which is then evaluated end-to-end by the evaluator. The evaluator returns key performance statistics, including the end-to-end makespan $T_{\mathrm{makespan}}$ and resource-utilization metrics. Guided by these feedback signals, the optimizer iteratively updates $\big(\{x_{k,f,t}\},\,\{y_{v,c}\},\,\{\pi_c\}\big)$ to improve the objective while satisfying system constraints, until a feasible, high-quality solution is obtained. The resulting policy is finally exported to the runtime execution stage and deployed as the RB allocation and mapping strategy, thereby closing the loop between optimization design and runtime execution.

\textit{2) Runtime Execution:} At runtime, modality slices are first handled by the communication-system module. Given the deployed RB allocation $\{x_{k,f,t}^{*}\}$, the communication module determines the slice release times $\{r_k\}$ and forwards the received slice payloads to the computation module. Upon release, the computation platform executes the corresponding modality-specific subgraphs according to the deployed mapping strategy $\big(\{y_{v,c}^{*}\},\,\{\pi_c^{*}\}\big)$. Execution terminates at the sink node $o$, producing the final inference outputs and the realized end-to-end makespan $T_{\mathrm{makespan}}$.

\section{Optimization Algorithms}\label{sec:algorithms}
The optimization problem (P1) couples OFDMA RB allocation with precedence-constrained parallel-machine scheduling, where the RB allocation determines the release times for DNN execution on the multi-core accelerator. Considering that (P1) is NP-hard and features a huge combinatorial design space, we develop two heuristic optimization algorithms\footnote{Within O-WiN, the optimization module is treated as an interface-defined, black-box component, allowing different scheduling strategies to be integrated interchangeably; any improved algorithm can replace our heuristic with minimal integration effort and without modifying other modules.}: (i) \emph{Release-Time First Scheduling (RTFS)}, a stage-wise transmit-then-compute heuristic that prioritizes uplink delivery to minimize release times before inference; and (ii) \emph{Pipeline-Aware Co-Scheduling (PACS)}, a goal-driven heuristic that interleaves communication and computation to minimize the makespan.

\subsection{RTFS Algorithm}\label{sec:RTFS}
In RTFS, communication and computation are decoupled into two sequential stages separated by a wait-all barrier: computation begins only after all slices have been received at the edge server. In the first stage, RB allocation is implemented to deliver all slices to the edge server. In the second stage, the DNN workload is executed by mapping jobs onto the multi-core accelerator, subject to both DAG precedence constraints and NoC bandwidth constraints.\par

\subsubsection{Communication Stage}
As DNN execution on the multi-core accelerator is deferred until all data slices arrive, the computation start time is determined by the latest slice arrival time:
\begin{equation}
t_{\mathrm{start}}=\max_{k\in\mathcal{K}}~r_k(\mathbf{x}) =\Delta\cdot \max_{k\in\mathcal{K}} t_k^{\mathrm{arr}}(\mathbf{x}).
\label{eq:ss_waitall}
\end{equation}
To reduce the wait-all latency, a tail-oriented objective is adopted to prioritize the slowest slice. Define the remaining payload of slice $k$ at the beginning of slot $t$ as
\begin{equation}
R_k(t;\mathbf{x})=\max\bigl\{0,\ D_k - B_k(t-1;\mathbf{x})\bigr\}.
\label{eq:ss_remaining}
\end{equation}
The residual transmission time of slice $k$ is given by
\begin{equation}
\hat{T}_k(t)=\frac{R_k(t;\mathbf{x})}{\hat{u}_k(t)},
\label{eq:ss_tailtime}
\end{equation}
where $\hat{u}_k(t)$ denotes the predicted effective uplink transmission rate of node $k$ during time slot $t$. 

The predicted effective uplink transmission rate $\hat{u}_k(t)$ is estimated in two steps based on future channel statistics. First, the instantaneous uplink transmission rate over all subcarriers is averaged to form the slot-level mean rate:
\begin{equation}
\bar{u}_k(\tau)=\frac{1}{F}\sum_{f=1}^{F}u_{k,f}(\tau),
\end{equation}
where $u_{k,f}(\tau)$ denotes the instantaneous uplink transmission rate of node $k$ on subcarrier $f$ during slot $\tau$. Next, the predicted effective uplink transmission rate at slot $t$ is defined as the suffix mean over the remaining horizon:
\begin{equation}
\hat{u}_k(t)=\frac{1}{T-t}\sum_{\tau=t}^{T-1}\bar{u}_k(\tau).
\end{equation}
For numerical stability, we apply a small floor $\epsilon>0$ with $\hat{u}_k(t)\leftarrow\max\{\hat{u}_k(t),\epsilon\}$. Within each slot $t$, subcarriers are sequentially assigned in a greedy manner to approximately minimize the maximum tail time $\max_{k\in\mathcal{K}}\hat{T}_k(t)$. For each subcarrier, a one-step update is performed by
\begin{equation}
R_k \leftarrow \max\bigl\{0,\ R_k - \Delta\,\rho_{k,f,t}\bigr\}.
\label{eq:ss_one_step}
\end{equation}
Subsequently, recompute $\max_{j\in\mathcal{K}}\hat{T}_j(t)$ and select the node that yields the smallest resulting maximum tail time. The communication stage of RTFS is summarized in Algorithm~\ref{alg:ss_comm_only}.

\algrenewcommand\algorithmicrequire{\textbf{Input:}}
\algrenewcommand\algorithmicensure{\textbf{Output:}}

\begin{algorithm}[t]
\caption{Communication Stage of RTFS}
\label{alg:ss_comm_only}
\begin{algorithmic}[1]
\Require $\mathcal{K}$, $\mathcal{F}$, $\Delta$, $\{D_k\}_{k\in\mathcal{K}}$, $\{\rho_{k,f,t}\}$, $\{\hat{u}_k(t)\}$, $\epsilon>0$
\Ensure $\{x_{k,f,t}\}$, $\{t_k^{\mathrm{arr}}\}$, $t_{\mathrm{start}}$

\State $R_k \gets D_k,\quad t_k^{\mathrm{arr}} \gets +\infty,\ \forall k\in\mathcal{K}$
\State $t \gets 0$
\While{$\exists k\in\mathcal{K}: R_k>0$}
    \State $\tilde{R}_k \gets R_k,\ \forall k\in\mathcal{K}$  
    \State $x_{k,f,t} \gets 0,\ \forall k\in\mathcal{K}, f\in\mathcal{F}$

    \For{$f\in\mathcal{F}$} 
        \State $k^\star \gets \bot,\quad \mathrm{best} \gets +\infty$
        \ForAll{$k\in\mathcal{K}$}
            \If{$\tilde{R}_k \le 0$}
                \State \textbf{continue}
            \EndIf
            \State $b \gets \Delta\,\rho_{k,f,t}$ 
            \State $\tilde{R}'_j \gets \tilde{R}_j,\ \forall j\in\mathcal{K}$  
            \State $\tilde{R}'_k \gets \max\{0,\ \tilde{R}_k - b\}$
            \State $\hat{T}_j \gets \tilde{R}'_j / \max\{\hat{u}_j(t),\epsilon\},\ \forall j\in\mathcal{K}$  
            \State $\mathrm{obj} \gets \max_{j\in\mathcal{K}} \hat{T}_j$  
            \If{$\mathrm{obj} < \mathrm{best}$}
                \State $\mathrm{best}\gets \mathrm{obj},\quad k^\star \gets k$
            \EndIf
        \EndFor
        \If{$k^\star \neq \bot$}
            \State $x_{k^\star,f,t} \gets 1$
            \State $\tilde{R}_{k^\star} \gets \max\{0,\ \tilde{R}_{k^\star}-\Delta\,\rho_{k^\star,f,t}\}$  
        \EndIf
    \EndFor

    \For{$f\in\mathcal{F}$}  
        \State Find $k$ such that $x_{k,f,t}=1$ (if none, continue)
        \State $R_k \gets \max\{0,\ R_k-\Delta\,\rho_{k,f,t}\}$
    \EndFor

    \ForAll{$k\in\mathcal{K}$}
        \If{$R_k=0$ \textbf{and} $t_k^{\mathrm{arr}}=+\infty$}
            \State $t_k^{\mathrm{arr}} \gets t$  
        \EndIf
    \EndFor
    \State $t \gets t+1$
\EndWhile
\State $t_{\mathrm{start}} \gets \Delta \cdot \max_{k\in\mathcal{K}} t_k^{\mathrm{arr}}$
\State \Return $\{x_{k,f,t}\}$, $\{t_k^{\mathrm{arr}}\}$, $t_{\mathrm{start}}$
\end{algorithmic}
\end{algorithm}

\subsubsection{Computation Stage}
In the computation stage, upon reaching the wait-all synchronization point at $t_{\mathrm{start}}$, all slice-dependent, modality-specific subgraphs are released concurrently, allowing inference to commence immediately at the edge server. Furthermore, DNN execution on the multi-core accelerator under a shared NoC bandwidth budget $B_{\max}$ is modeled as an event-driven simulator.

A job becomes ready only when all its predecessors have finished.  At each decision point, ready jobs are dispatched to idle cores using a rank-based list scheduling policy in the spirit of heterogeneous earliest finish time (HEFT). The upward rank of  $v\in \mathcal{V}$ is given by
\begin{equation}
\mathrm{rank}_u(v) \triangleq \overline{L}_v + \max_{k\in \mathrm{succ}(v)} \mathrm{rank}_u(k),
\label{eq:ttc_ranku}
\end{equation}
where $\overline{L}_v=\frac{1}{C}\sum_{c=1}^C L_{v,c}$ denotes the average execution latency of job $v$ across all cores, and $\mathrm{succ}(v)$ denotes the successor set of $v$ in the DAG; for exit jobs with no successors, the maximization term is defined to be zero. Intuitively, $\mathrm{rank}_u(v)$ serves as a proxy for the remaining critical-path cost starting from job $v$, and thus provides a priority metric for online scheduling. At runtime, whenever a core becomes idle, the scheduler constructs the ready set consisting of jobs whose predecessors have completed and whose required inputs are available. The scheduler then selects a job with the largest upward rank $\mathrm{rank}_u(\cdot)$ from the ready set and dispatches it to the idle core for non-preemptive execution until completion; ties may be broken by a secondary criterion such as smaller $\overline{L}_v$.

The simulator advances time directly to the next completion event, updates remaining work for all running jobs, marks completions, and releases newly ready successors. Given the DAG $\mathcal{G}=(\mathcal{V},\mathcal{E})$, jobs are dispatched online onto $C$ accelerator cores, where a job becomes eligible once all its predecessors have completed. In addition, execution is subject to contention on the NoC bandwidth: when multiple cores run concurrently, their jobs compete for a shared bandwidth budget, which affects job progress and hence the completion time of each job. Specifically, each job has a bandwidth requirement $b_{v,c}\in(0,1]$. The platform provides a global shared resource budget $B_\text{max}$. When multiple cores execute jobs concurrently, the actual bandwidth allocated to a running job on core $c$ is proportional to its requested $b_{v_c,c}$:
\begin{equation}
\alpha_c(t)=
\begin{cases}
\dfrac{b_{v_c,c}}{\sum_{k\in \mathcal{R}(t)} b_{v_k,k}}\cdot B_\text{max}, & \mathcal{R}(t)\neq\varnothing,\\[8pt]
0, & \text{otherwise},
\end{cases}
\label{eq:bwshare}
\end{equation}
where $\mathcal{R}(t)$ is the set of cores running jobs at time $t$, and $v_k$ denotes the job currently assigned to core $k$.

\subsection{PACS Algorithm}\label{subsec:PACS}
In RTFS, the wait-all barrier often leaves compute resources idle, thereby potentially prolonging the end-to-end makespan. In contrast, PACS removes this barrier and releases eligible slice-dependent jobs immediately upon slice arrival, enabling pipelined communication–computation overlap and more effectively exploiting cross-modality concurrency. \par

\subsubsection{Communication Stage}
Unlike RTFS, PACS evaluates each candidate RB allocation decision based on its downstream effect on DNN execution on the multi-core accelerator, explicitly coupling radio resource allocation with accelerator-side execution.
However, estimating the exact makespan for each candidate by fully re-simulating the communication–computation pipeline is prohibitively expensive for online decision making. Given this, a lightweight predictor $\mathcal{P}$ is employed to approximate the completion time of the final sink node. Instead of running a full multi-core execution simulation, the predictor couples communication-induced slice arrivals with a coarse execution model parameterized by offline-profiled average operator latencies. In particular, core contention and queueing effects are intentionally ignored to keep the prediction overhead low.\par

At time slot $t$, the predictor $\mathcal{P}$ estimates the arrival time of slice $k$ as
\begin{equation}
\hat{A}_k(t)=
\begin{cases}
t, & R_k(t)\le 0,\\[4pt]
t + \dfrac{R_k(t)}{\hat{u}_k(t)}, & R_k(t)>0.
\end{cases}
\label{eq:joint_arrival_est_fix}
\end{equation}
The corresponding release time for the corresponding entry job is set to
\begin{equation}
S_v(t)=\max\{t,\hat{A}_k(t)\},\qquad \forall v\in\mathcal{V}_k.
\label{eq:release_time}
\end{equation}

\algrenewcommand\algorithmicrequire{\textbf{Input:}}
\algrenewcommand\algorithmicensure{\textbf{Output:}}

\begin{algorithm}[H]
\caption{Computation Stage of RTFS}
\label{alg:compute_engine}
\small
\begin{algorithmic}[1]
\Require $\mathcal{V}$, $L_{v,c}$ and $b_{v,c}$ for $c\in \mathcal{C}$, $B_{\max}$, $\epsilon>0$, $t_{\mathrm{start}}$;
$\mathrm{op}(v)$, $\mathrm{slice}(v)$, and $\{g_k\}_{k=1}^{K}$.
\Ensure $\{s_v\}_{v\in\mathcal{V}}$, $\{f_v\}_{v\in\mathcal{V}}$

\For{$v\in\mathcal{V}$}
    \State $\overline{L}_v \gets \frac{1}{C}\sum_{c=1}^{C} L_{v,c}$
\EndFor
\ForAll{$v\in\mathcal{V}$ in reverse topo order}
    \If{$\mathrm{succ}(v)=\varnothing$}
        \State $\mathrm{rank}_u(v)\gets \overline{L}_v$
    \Else
        \State $\mathrm{rank}_u(v)\gets \overline{L}_v + \max_{k\in\mathrm{succ}(v)} \mathrm{rank}_u(k)$
    \EndIf
\EndFor

\vspace{2pt}
\Statex \textit{(1) Initialization}
\State $t_{\mathrm{cur}} \gets t_{\mathrm{start}}$
\State $\sigma(v)\gets 0,\ \forall v\in\mathcal{V}$ 
\State $s_v \gets \bot,\ f_v\gets \bot,\ \forall v$
\State $\mathcal{Q}_{\mathrm{ready}}\gets \varnothing$

\ForAll{$v\in\mathcal{V}$}
    \If{$\mathrm{pred}(v)=\varnothing$ \textbf{and} $\bigl(\mathrm{op}(v)\neq \texttt{Embed}\ \textbf{or}\ t_{\mathrm{cur}}+\epsilon \ge g_{\mathrm{slice}(v)}\bigr)$}
        \State $\sigma(v)\gets 1$
        \State $\mathcal{Q}_{\mathrm{ready}}\gets \mathcal{Q}_{\mathrm{ready}}\cup\{v\}$
    \EndIf
\EndFor
\State $v_c \gets \text{null},\ \omega_c \gets 0,\ \forall c\in \mathcal{C}$

\vspace{2pt}
\While{$\exists v\in\mathcal{V}: \sigma(v)\neq 3$}

    \Statex \textit{(2) Dispatch ready jobs to idle cores}
    \State $\mathcal{I}\gets \{c: v_c=\text{null}\}$
    \While{$\mathcal{I}\neq \varnothing$ \textbf{and} $\mathcal{Q}_{\mathrm{ready}}\neq \varnothing$}
        \State Choose any $c^\star \in \mathcal{I}$
        \State $v^\star \gets \arg\min\limits_{v\in\mathcal{Q}_{\mathrm{ready}}}\ \Bigl(-\mathrm{rank}_u(v),\ \overline{L}_v\Bigr)$
        \If{$s_{v^\star} = \bot$}
            \State $s_{v^\star} \gets t_{\mathrm{cur}}$
        \EndIf
        \State $v_{c^\star}\gets v^\star$
        \State $\omega_{c^\star}\gets L_{v^\star,c^\star}\cdot b_{v^\star,c^\star}$ 
        \State $\sigma(j^\star)\gets 2$
        \State $\mathcal{Q}_{\mathrm{ready}}\gets \mathcal{Q}_{\mathrm{ready}}\setminus\{v^\star\}$
        \State $\mathcal{I}\gets \mathcal{I}\setminus\{c^\star\}$
    \EndWhile

    \Statex \textit{(3) Event-driven progress under bandwidth sharing}
    \State $\mathcal{C}_{\mathrm{run}}\gets \{c: v_c\neq\text{null}\}$

    \State $Z \gets \sum\limits_{c\in\mathcal{C}_{\mathrm{run}}} b_{v_c,c}$
    \If{$Z>\epsilon$}
        \ForAll{$c\in\mathcal{C}_{\mathrm{run}}$}
            \State $\alpha_c \gets \frac{b_{v_c,c}}{Z}\cdot B_{\max}$ 
        \EndFor
    \Else
        \ForAll{$c\in\mathcal{C}_{\mathrm{run}}$}
            \State $\alpha_c \gets \frac{B_{\max}}{|\mathcal{C}_{\mathrm{run}}|}$
        \EndFor
    \EndIf

    \State $\delta \gets \min\limits_{c\in\mathcal{C}_{\mathrm{run}}}\ \omega_c/(\alpha_c+\epsilon)$
    \State $t_{\mathrm{cur}} \gets t_{\mathrm{cur}}+\delta$
    \ForAll{$c\in\mathcal{C}_{\mathrm{run}}$}
        \State $\omega_c \gets \omega_c - \alpha_c\cdot \delta$
    \EndFor

    \Statex \textit{(4) Complete jobs and update ready set}
    \ForAll{$c\in\mathcal{C}_{\mathrm{run}}$}
        \If{$\omega_c \le \epsilon$}
            \State $v\gets v_c$
            \State $f_v \gets t_{\mathrm{cur}}$
            \State $\sigma(v)\gets 3$ 
            \State $v_c\gets \text{null},\ \omega_c\gets 0$
        \EndIf
    \EndFor

    \ForAll{$v\in\mathcal{V}$}
        \If{$\sigma(v)=0$ \textbf{and} $\mathrm{pred}(v)\subseteq \{k\in\mathcal{V}:\sigma(k)=3\}$ \textbf{and}
        $\bigl(\mathrm{op}(v)\neq \texttt{Embed}\ \textbf{or}\ t_{\mathrm{cur}}+\epsilon \ge g_{\mathrm{slice}(v)}\bigr)$}
            \State $\sigma(v) \gets 1$
            \State $\mathcal{Q}_{\mathrm{ready}}\gets \mathcal{Q}_{\mathrm{ready}}\cup\{v\}$
        \EndIf
    \EndFor

\EndWhile

\State \Return $\{s_v\}_{v\in\mathcal{V}}$, $\{f_v\}_{v\in\mathcal{V}}$
\end{algorithmic}
\end{algorithm}

To capture time-varying execution rates under bandwidth sharing, we define the \emph{work} of executing job $v$ on core $c$ as
\begin{equation}
W_{v,c}=L_{v,c}\cdot b_{v,c},
\label{eq:work}
\end{equation}
where $L_{v,c}$ is the base latency of job $v$ on core $c$. During execution, the remaining work of the running job on core $c$ decreases at rate $\alpha_c(t)$, which naturally yields piecewise progress as the set of concurrently running jobs changes over time. The computation stage of RTFS is summarized in Algorithm~\ref{alg:compute_engine}.

Given these release-time constraints, the earliest finish time of job $v$ is predicted via a max-plus dynamic program on the DAG, which is given by
\begin{equation}
\hat{F}_v(t)=
\begin{cases}
S_v(t)+\bar{L}_v, & \mathrm{pred}(v)=\varnothing,\\[6pt]
\bar{L}_v+\max\limits_{v' \in \mathrm{pred}(v)} \hat{F}_{v'}(t), & \mathrm{pred}(v)\neq\varnothing.
\end{cases}
\label{eq:dag_dp_general}
\end{equation}
This recursion naturally captures an arbitrary number of branches and synchronization points: any multi-input operator is simply a job with $|\mathrm{pred}(v)|>1$, whose start time is determined by the latest predecessor completion. The predicted makespan from the current remaining payloads is given by
\begin{equation}
\hat{T}_{\mathrm{makespan}}(t)=\hat{F}_o(t).
\label{eq:makespan_proxy_general}
\end{equation}
\par Upon completion of the makespan estimation, subcarriers are allocated greedily within each time slot: each subcarrier is assigned to the data slice that achieves the largest marginal reduction in the predicted output completion time.
For unfinished data slice candidate $k$ with $R_k(t;\mathbf{x})>0$, a trial remaining vector $R^{(k)}$ is constructed by reducing only slice $k$ by the amount deliverable on subcarrier $f$, which is given by
\begin{equation}
R^{(k)}_{k'} \leftarrow
\begin{cases}
\max\bigl\{0,\ R_k(t)-\Delta\,\rho_{k,f,t}\bigr\}, & k'=k,\\[4pt]
R_{k'}(t), & k'\neq k.
\end{cases}
\label{eq:joint_trial_unified_case}
\end{equation}
Then, the corresponding marginal decrease is given by
\begin{equation}
G(k) = \hat{T}_{\mathrm{makespan}}(t; R_k) - \hat{T}_{\mathrm{makespan}}(t; R^{(k)}).
\label{eq:joint_gain}
\end{equation}
Based on this, subcarrier $f$ is assigned to the slice $k^\star=\arg\max_k G(k)$, after which the remaining vector is updated as $R_k \leftarrow R^{(k^\star)}$ and the procedure continues to subcarrier $f{+}1$. By directly maximizing the reduction in $\hat{T}_{\mathrm{makespan}}$, this greedy allocation prioritizes slices that are most critical to imminent synchronization points and the end-to-end critical path to output. The communication stage of PACS is summarized in Algorithm~\ref{alg:pacs_comm_full}.

\algrenewcommand\algorithmicrequire{\textbf{Input:}}
\algrenewcommand\algorithmicensure{\textbf{Output:}}

\begin{algorithm}[t]
\caption{Communication Stage of PACS}
\label{alg:pacs_comm_full}
\begin{algorithmic}[1]
\Require $\mathcal{K}$, $\mathcal{F}$, $\Delta$, $\{D_k\}_{k\in\mathcal{K}}$, $\{\rho_{k,f,t}\}$, $\{\hat{u}_k(t)\}$, $\mathcal{P}$, $\epsilon>0$
\Ensure $\{x_{k,f,t}\}$, $\{t_k^{\mathrm{arr}}\}$

\State $R_k \gets D_k,\quad t_k^{\mathrm{arr}} \gets +\infty,\ \forall k\in\mathcal{K}$
\State $t \gets 0$
\While{$\exists k\in\mathcal{K}: R_k>0$}
    \State $\tilde{R}_k \gets R_k,\ \forall k\in\mathcal{K}$ 
    \State $x_{k,f,t} \gets 0,\ \forall k\in\mathcal{K}, f\in\mathcal{F}$
    
    \State $\hat{T}_{\mathrm{base}} \gets \mathcal{P}\!\left(t,\ \tilde{R},\ \hat{u}(t)\right)$

    \For{$f\in\mathcal{F}$}
        \State $k^\star \gets \bot,\quad G_\text{best} \gets -\infty$
        \ForAll{$k\in\mathcal{K}$}
            \If{$\tilde{R}_k \le 0$}
                \State \textbf{continue}
            \EndIf
            \State $b \gets \Delta \cdot \rho_{k,f,t}$
            \State $\tilde{R}'_j \gets \tilde{R}_j,\ \forall j\in\mathcal{K}$
            \State $\tilde{R}'_k \gets \max\{0,\ \tilde{R}_k - b\}$
            
            \State $\widehat{T}_{\mathrm{try}} \gets \mathcal{P}\!\left(t,\ \tilde{R}',\ \hat{u}(t)\right)$
            \State $G \gets \hat{T}_{\mathrm{base}} - \hat{T}_{\mathrm{try}}$

            \If{$G > G_\text{best}$}
                \State $G_\text{best}\gets G,\quad k^\star \gets k$
            \EndIf
        \EndFor

        \If{$k^\star \neq \bot$}
            \State $x_{k^\star,f,t} \gets 1$
            \State $\tilde{R}_{k^\star} \gets \max\{0,\ \tilde{R}_{k^\star} - \Delta\cdot\rho_{k^\star,f,t}\}$
            \State $\hat{T}_{\mathrm{base}} \gets \mathcal{P}\!\left(t,\ \tilde{R},\ \hat{u}(t)\right)$
        \EndIf
    \EndFor

    \For{$f\in\mathcal{F}$}
        \State Find $k$ such that $x_{k,f,t}=1$ (if none, continue)
        \State $R_k \gets \max\{0,\ R_k - \Delta\cdot\rho_{k,f,t}\}$
    \EndFor

    \ForAll{$k\in\mathcal{K}$}
        \If{$R_k = 0$ \textbf{and} $t_k^{\mathrm{arr}}=+\infty$}
            \State $t_k^{\mathrm{arr}} \gets t$
        \EndIf
    \EndFor

    \State $t \gets t+1$
\EndWhile

\State \Return $\{x_{k,f,t}\}$, $\{t_k^{\mathrm{arr}}\}$
\end{algorithmic}
\end{algorithm}
\subsubsection{Computation Stage}
The computation scheduler follows the same rank-based list scheduling policy as in RTFS, dispatching the highest-priority ready jobs onto idle cores. The difference is that the ready set evolves jointly with communication: as soon as a slice arrives, its gate time $A_k$ is reached and the corresponding slice-gated jobs can enter the ready set, allowing computation to overlap with ongoing transmission. Slice arrival induces an input gate: the first job of slice $k$ can become ready only when $t \ge A_k$, while all other jobs follow standard DAG precedence constraints. The system advances in an event-driven manner to the earliest of (i) the next computation completion event and (ii) the next communication tick boundary, and then updates the ready set and dispatch decisions accordingly. The computation stage of PACS is summarized in Algorithm~\ref{alg:compute_engine_comm_input}.

\algrenewcommand\algorithmicrequire{\textbf{Input:}}
\algrenewcommand\algorithmicensure{\textbf{Output:}}

\begin{algorithm}[t]
\caption{Computation Stage of RTFS}
\label{alg:compute_engine_comm_input}
\small
\begin{algorithmic}[1]
\Require $\mathcal{V}$, $L_{v,c}$ and $b_{v,c}$ for $c\in \mathcal{C}$, $B_{\max}$, $\epsilon>0$, $t_{\mathrm{start}}$;
$\mathrm{op}(v)$, $\mathrm{slice}(v)$; $\{t_k^{\mathrm{arr}}\}_{k=1}^{K}$.
\Ensure $\{s_v\}_{v\in\mathcal{V}}$, $\{f_v\}_{v\in\mathcal{V}}$

\State Algorithm 2, lines 1–7

\vspace{2pt}
\Statex \textit{(1) Initialization}
\State $t_{\mathrm{cur}} \gets t_{\mathrm{start}}$
\State $g_k \gets t_k^{\mathrm{arr}},\ \forall k$ 

\State Algorithm 2, lines 9–16

\vspace{2pt}
\While{$\exists v\in\mathcal{V}: \sigma(v)\neq 3$}

    \Statex \textit{(2) Dispatch ready jobs to idle cores}
    \State Algorithm 2, lines 18–28

    \Statex \textit{(3) Event-driven progress under bandwidth sharing}
    \State $\mathcal{C}_{\mathrm{run}}\gets \{c: v_c\neq\text{null}\}$

    \If{$\mathcal{C}_{\mathrm{run}}=\varnothing$}
        \State $t_{\mathrm{comp}} \gets +\infty$
    \Else
        \State $Z \gets \sum\limits_{c\in\mathcal{C}_{\mathrm{run}}} b_{v_c,c}$
        \If{$Z>\epsilon$}
            \ForAll{$c\in\mathcal{C}_{\mathrm{run}}$}
                \State $\alpha_c \gets \frac{b_{v_c,c}}{Z}\cdot B_{\max}$
            \EndFor
        \Else
            \ForAll{$c\in\mathcal{C}_{\mathrm{run}}$}
                \State $\alpha_c \gets \frac{B_{\max}}{|\mathcal{C}_{\mathrm{run}}|}$
            \EndFor
        \EndIf
        \State $\delta_{\mathrm{comp}} \gets \min\limits_{c\in\mathcal{C}_{\mathrm{run}}}\ \omega_c/(\alpha_c+\epsilon)$
        \State $t_{\mathrm{comp}} \gets t_{\mathrm{cur}}+\delta_{\mathrm{comp}}$
    \EndIf

    \State $t_{\mathrm{gate}} \gets \min\{g_k: g_k>t_{\mathrm{cur}}+\epsilon\}$
    \If{no such $k$ exists} \State $t_{\mathrm{gate}}\gets +\infty$ \EndIf

    \State $t_{\mathrm{next}}\gets \min\{t_{\mathrm{comp}},\ t_{\mathrm{gate}}\}$
    \If{$t_{\mathrm{next}}=+\infty$} \State \textbf{break} \EndIf
    \State $\delta \gets t_{\mathrm{next}}-t_{\mathrm{cur}}$
    \If{$\delta>0$ \textbf{and} $\mathcal{C}_{\mathrm{run}}\neq\varnothing$}
        \ForAll{$c\in\mathcal{C}_{\mathrm{run}}$}
            \State $\omega_c \gets \omega_c - \alpha_c\cdot \delta$
        \EndFor
    \EndIf
    \State $t_{\mathrm{cur}} \gets t_{\mathrm{next}}$

    \Statex \textit{(4) Complete jobs and update ready set}
    \State Algorithm 2, lines 41–50

\EndWhile

\State \Return $\{s_v\}_{v\in\mathcal{V}}$, $\{f_v\}_{v\in\mathcal{V}}$
\end{algorithmic}
\end{algorithm}

\subsection{Complexity Analysis}\label{subsec:complexity}
Let $V\triangleq|\mathcal{V}|$ and $E\triangleq|\mathcal{E}|$ denote the numbers of jobs and precedence edges in the multimodal DAG, respectively. For both RTFS and PACS, the upward ranks $\{\mathrm{rank}_u(v)\}$ are precomputed once via a reverse-topological dynamic programming (DP) in $\mathcal{O}(V+E)$.

\subsubsection{RTFS}
In each slot, RTFS assigns each of the $F$ subcarriers to one of the $K$ slices using a greedy search that evaluates up to $K$ candidates per subcarrier and updates the predicted remaining times for up to $K$ slices, yielding $\mathcal{O}(F K^{2})$ per slot and $\mathcal{O}(T F K^{2})$ over $T$ slots. After all slices arrive, RTFS runs an event-driven list scheduler: each job is dispatched and completed once; each completion triggers bandwidth re-allocation over at most $C$ active cores and updates the ready set. With a priority queue keyed by $\mathrm{rank}_u$, the scheduling overhead is $\mathcal{O}(V\log V)$ and event processing contributes $\mathcal{O}(VC)$, giving $\mathcal{O}(V\log V + VC)$ for computation (excluding rank precomputation).

\subsubsection{PACS}
PACS uses a lightweight predictor $\mathcal{P}$ that computes the output completion time via a max-plus DP on the DAG; given release times, $\mathcal{P}$ runs in $\mathcal{O}(V+E)$. In each slot, PACS assigns each subcarrier by enumerating up to $K$ candidate slices and invoking $\mathcal{P}$ to evaluate the marginal gain, resulting in $\mathcal{O}(F K (V+E))$ per slot and $\mathcal{O}(T F K (V+E))$ overall. PACS adopts the same event-driven list scheduler as RTFS, with the ready set additionally gated by slice arrival times, so the computation-stage complexity remains $\mathcal{O}(V\log V + VC)$.

\section{Evaluation}\label{sec:evaluation}
This section develops an event-driven simulator that co-models wireless transmission and precedence-constrained multi-core accelerator execution, enabling end-to-end evaluation of multimodal DNN workloads in WNP\footnote{These synthetic inputs can be seamlessly replaced with real measurements. Additionally, simulation enables controlled exploration of a broader parameter space while mitigating device- and environment-specific sampling bias, thereby improving generality and reproducibility.}. \par

\subsection{Simulation Setup}\label{sec:setup}
\begin{figure}
\centering
\includegraphics[width=3.2in,trim=0cm 0cm 0cm 0cm,clip]{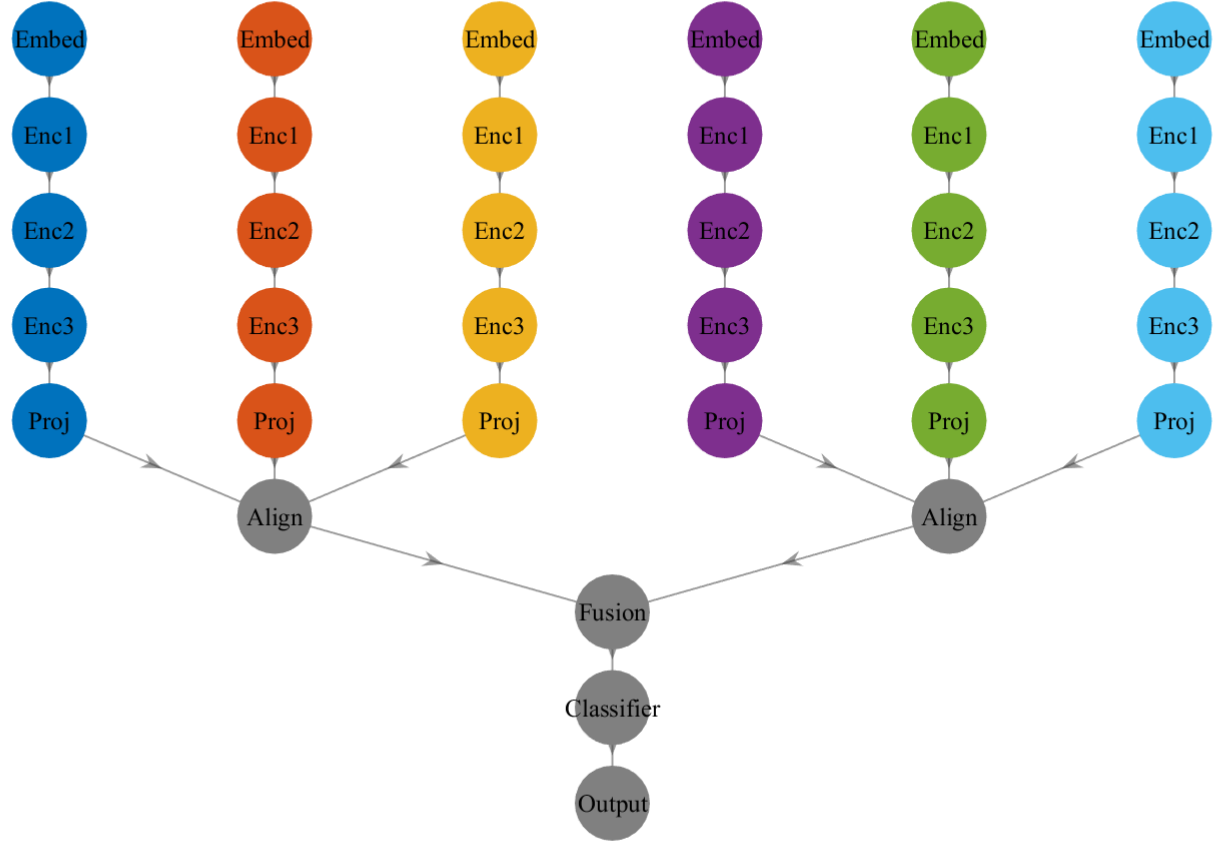}
\caption{Network architecture of multimodal DNN workload.}
\label{Fig3}
\end{figure}
For the multimodal DNN workload, a two-level DAG is constructed to capture both modality-parallel execution and cross-modality synchronization, as illustrated in Fig.~\ref{Fig3}. Specifically, the input is partitioned into $K$ modalities. For each modality $k$, a modality-specific subgraph comprises five sequential operators: $\texttt{Embed}_k \rightarrow \texttt{Enc1}_k \rightarrow \texttt{Enc2}_k \rightarrow \texttt{Enc3}_k \rightarrow \texttt{Proj}_k$.
The per-modality outputs are then synchronized and aggregated by a cross-modality subgraph. In particular, two alignment nodes wait for disjoint halves of the modalities:
$\texttt{Align}_1,\,\texttt{Align}_2 \rightarrow \texttt{Fusion} \rightarrow \texttt{Classifier} \rightarrow \texttt{Output}$, where $\texttt{Align}_1$ depends on $\{\texttt{Proj}_k\}_{k=0}^{\lfloor K/2 \rfloor-1}$ and $\texttt{Align}_2$ depends on $\{\texttt{Proj}_k\}_{k=\lfloor K/2 \rfloor}^{K-1}$. This fan-out--fan-in structure introduces multi-input synchronization points and is straggler-sensitive, thereby revealing communication--computation pipelining opportunities and making the end-to-end makespan jointly determined by transmission completion and compute scheduling.

For wireless transmission, per-slot per-subcarrier achievable rates $\rho_{k,f,t}$ are generated as independent and identically distributed (i.i.d.) samples within a range to simulate time-varying channels, and a suffix-mean predictor is used to estimate future effective rates. On the computing side, each job $v\in\mathcal{V}$ has a core-dependent latency $L_{v,c}$ and a bandwidth demand $b_{v,c}$, both synthetically generated as functions of the slice token size and the operator type. The multi-core accelerator enforces a shared NoC bandwidth budget $B_{\max}$; at runtime, bandwidth is dynamically shared among running cores in proportion to their demands, and each job is advanced in an event-driven manner based on its remaining bandwidth-work. Table~\ref{tab:sim_parameters} summarizes the key parameters used in our experiments.\par

Let $N_k$ denote the per-slice token budget, which can be used as the unified scale factor for both communication and computation. The communication payload of slice $k$ is given by
\begin{equation}
D_k =N_k \, D_\text{f} \, b \, \kappa_k \, \eta,
\label{eq:payload_short}
\end{equation}
where $D_\text{f}$ is the feature dimension, $b$ is the bytes per element, $\kappa_k$ is a slice-specific compression factor, and $\eta$ is the protocol overhead. For job $v$, the token scale is
\begin{equation}
S(v)=
\begin{cases}
N_k, & \text{if } v \text{ belongs to slice } k,\\
\sum_{k=1}^{K}N_k, & \text{if } v \text{ is cross--modality,}
\end{cases}
\label{eq:tokenscale_short}
\end{equation}
and the per-core job latency is generated by an operator-specific model with heterogeneity and jitter:
\begin{equation}
\ell_{v,c}=g_{\mathrm{op}(v)}\!\left(S(v)\right)\cdot \frac{\xi_{v,c}}{v_c}\cdot \Delta.
\label{eq:latency_short}
\end{equation}
Here, $g_{\mathrm{op}(v)}(\cdot)$ denotes the operator-dependent baseline latency model (e.g., linear for embedding/projection and including a quadratic term for encoder blocks), $\xi_{v,c}\in[0.9,1.1]$ is a multiplicative jitter factor capturing run-to-run variability, $v_c$ is the relative speed of core $c$ modeling heterogeneous compute, where $v_c = \max(1+u_c, 0.7)$ and $u_c \sim \mathcal{U}(-0.08, 0.08)$. Thus, increasing $N_k$ simultaneously increases slice payload $D_k$ and the execution times of all jobs whose workload depends on slice $k$.\par

\begin{table}[t]
\centering
\caption{Simulation Parameters}
\label{tab:sim_parameters}
\scriptsize
\setlength{\tabcolsep}{4pt}
\renewcommand{\arraystretch}{1.05}
\begin{tabular}{lcl}
\toprule
\textbf{Parameter} & \textbf{Symbol} & \textbf{Value / Range} \\ 
\midrule
Number of nodes/modalities  & $K$ & 6 \\
Slice token sizes & --- & $[512,128,256,128,192,128]$ \\
Payload model & --- & $D{=}512$, fp16 ($2$ bytes), overhead $1.05$ \\
Compression factors & --- & $[0.25,0.25,0.80,0.60,0.60,0.60]$ \\
Slice sizes & $D_k$ & $[134.4,33.6,215.0,80.6,121.0,80.6]$ kB \\
Accelerator cores & $C$ & 4 \\
NoC bandwidth budget & $B_{\max}$ & 1.0 (normalized) \\
Communication step & $\Delta$ & 1 ms \\
Subcarriers per slot & $F$ & 16 \\
Rate range (per subcarrier) & $\rho_{k,f,t}$ & 1--10 Mbps (1000--10000 kbps) \\
Communication trace length & $T$ & 8000 time slots \\
Core speed jitter & --- & $\pm 0.08$ \\
Latency scaling & --- & $\times 6$ \\
\bottomrule
\end{tabular}
\end{table}

\subsection{End-to-End Inference Latency}
\begin{figure*}[t]
\centering
\includegraphics[width=7.2in,trim=0cm 0cm 0cm 0cm,clip]{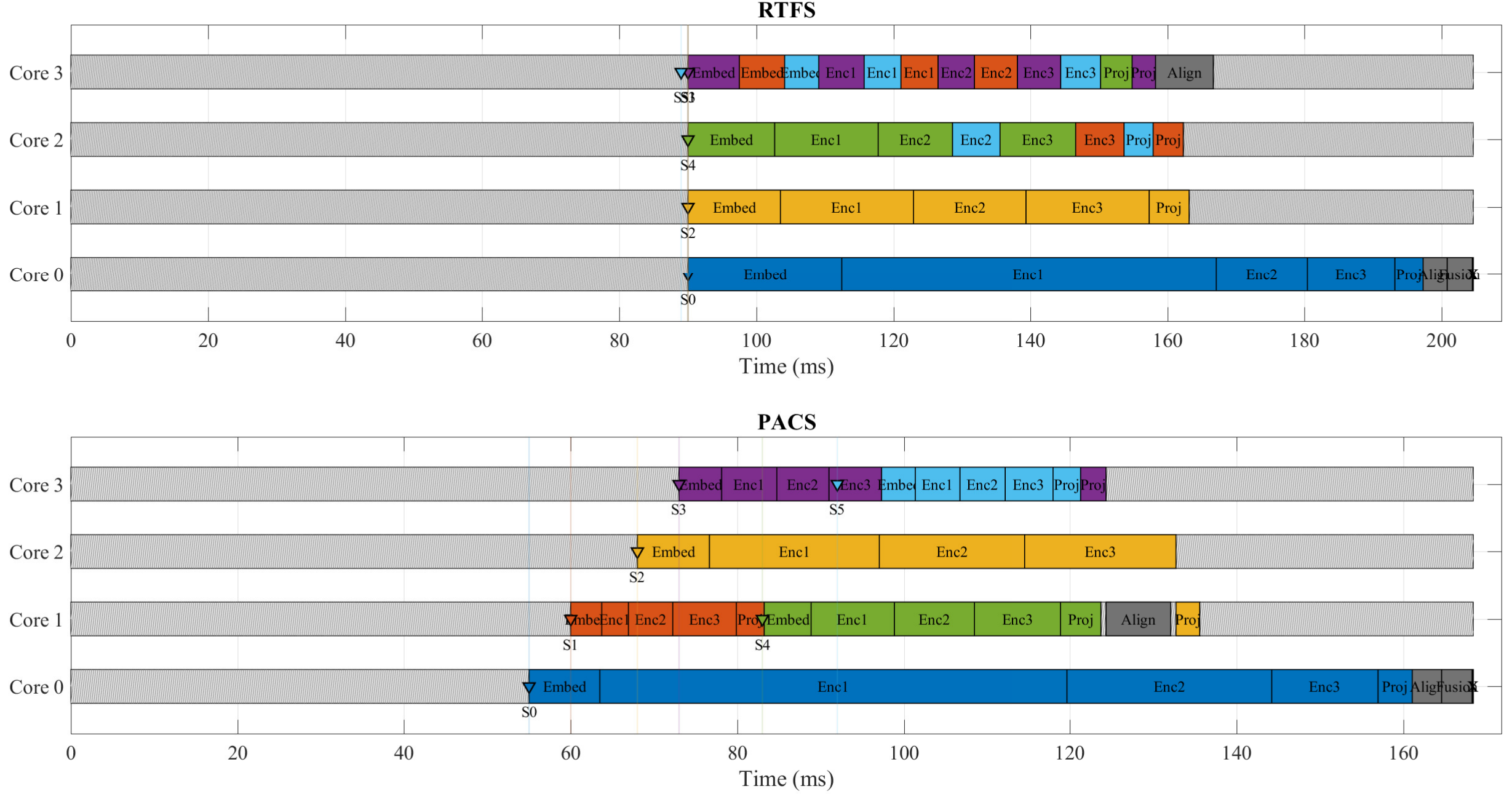}
\caption{The end-to-end execution timelines of RTFS and PACS. } 
\label{Fig4}
\end{figure*}

Fig.~\ref{Fig4} compares the end-to-end execution timelines of RTFS and PACS. Under RTFS, computation is separated from communication by a wait-all barrier: the accelerator cannot begin executing the entry \texttt{Embed} jobs until the modality slice has arrived. Consequently, the wireless tail latency is exposed as a pure idle interval on the compute side and is added almost directly to the end-to-end makespan. By contrast, PACS eliminates the barrier and activates modality-specific subgraphs in a slice-gated manner. As soon as a slice becomes available, its encoder jobs are released and can run in parallel with ongoing uplink transmission of the remaining modalities. This communication--computation overlap substantially improves core utilization, reduces idle gaps, and advances the completion of the terminal \texttt{Output} node. Furthermore, PACS allocates RBs based on the marginal reduction in the predicted \texttt{Output} completion time, thereby prioritizing slices that lie on (or gate) the cross-modality synchronization path (e.g., \texttt{Align} and \texttt{Fusion}) and yielding a smaller end-to-end makespan.

\begin{figure*}[t]
\centering
\includegraphics[width=7.2in,trim=0cm 0cm 0cm 0cm,clip]{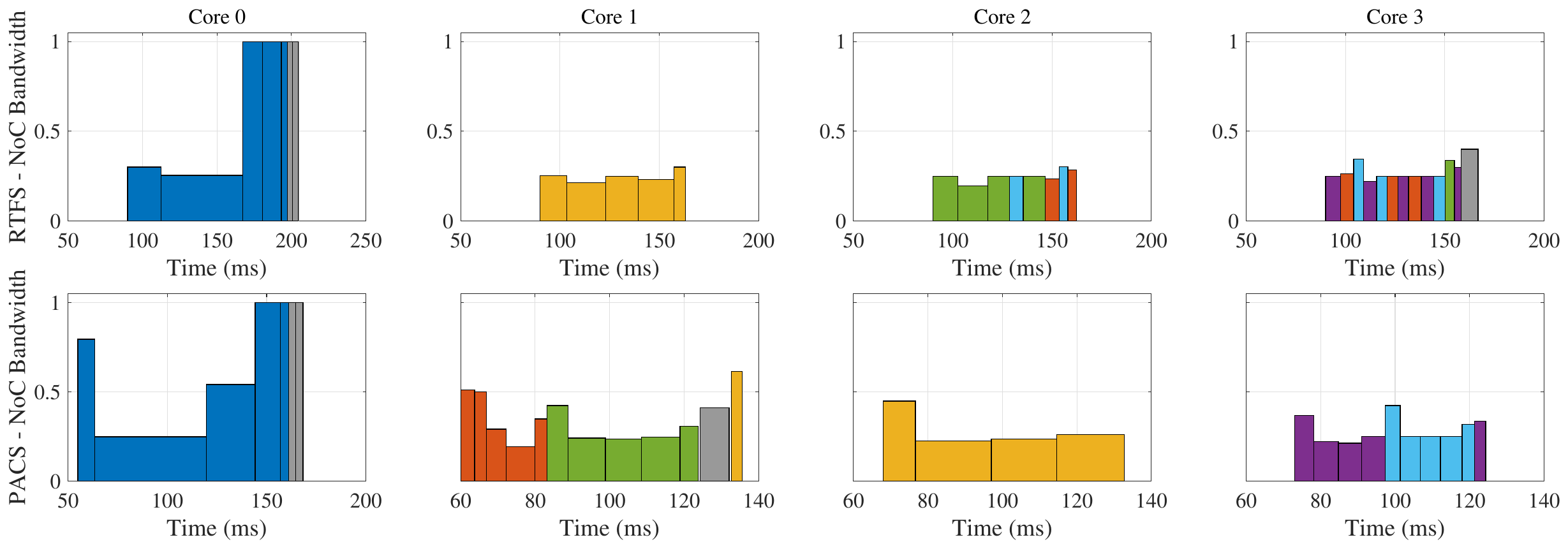}
\caption{NoC bandwidth sharing on the multi-core accelerator of RTFS and PACS.}
\label{Fig5}
\end{figure*}
Fig.~\ref{Fig5} further contrasts RTFS and PACS from the perspective of NoC bandwidth sharing on the multi-core accelerator. The plotted bars represent the instantaneous per-core shares of the normalized NoC budget, with colors indicating slice-specific jobs. Compared with RTFS, PACS exhibits more sustained bandwidth utilization with fewer prolonged low-activity intervals, consistent with the reduced idle gaps and the improved communication--computation overlap observed in Fig.~\ref{Fig4}. Moreover, because slices from different modalities arrive in a staggered fashion, NoC bandwidth contention is alleviated, which further speeds up computational throughput and overall execution.\par
\begin{figure*}[t]
\centering
\includegraphics[width=7.2in,trim=0cm 0cm 0cm 0cm,clip]{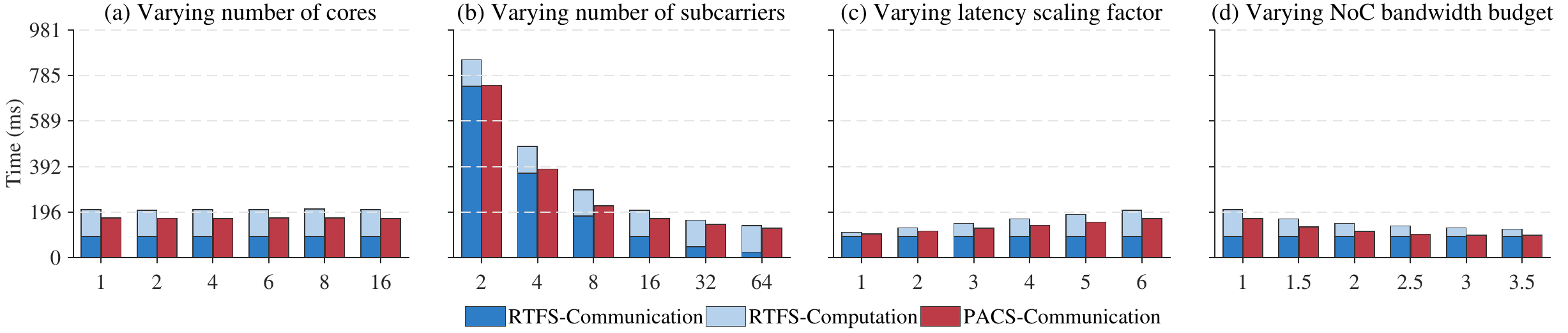}
\caption{The effects of four system-scaling dimensions—(a) the number of accelerator cores, (b) the number of OFDMA subcarriers, (c) the wireless-latency scaling factor, and (d) the NoC bandwidth budget—on the end-to-end makespan of RTFS and PACS.}
\label{Fig6}
\end{figure*}
Fig.~\ref{Fig6} shows an end-to-end makespan breakdown for RTFS and PACS under four system-scaling dimensions: (a) varying the number of accelerator cores, (b) varying the number of OFDMA subcarriers, (c) varying the wireless latency scaling factor, and (d) varying the NoC bandwidth budget. For RTFS, the makespan is explicitly decomposed into communication and computation components as stacked bars, reflecting its wait-all execution that largely serializes the two phases. In contrast, PACS enables slice-gated execution and overlaps computation with ongoing transmission. Among these subplots, the performance trend is most pronounced when the number of subcarriers (i.e., the communication resource) changes, whereas variations in the compute-related parameters induce only minor shifts. This contrast indicates that the overall makespan is primarily communication-constrained.

Table~\ref{tab:main_with_vectors} reports results under different token lengths and compression settings. The data suggest that when token lengths and compression ratios are relatively balanced across modalities—so that their communication and computation loads are comparable—the advantage of RTFS over PACS is not pronounced. In some cases, PACS can even perform slightly worse than RTFS. In contrast, when token lengths and compression ratios differ substantially across modalities—leading to highly imbalanced communication and computation demands—PACS exhibits a clear advantage over RTFS. Collectively, these results show that when modality heterogeneity is high, PACS can more effectively hide wireless latency behind computation.

\begin{table*}[t]
\centering
\caption{The end-to-end makespan $T_{\mathrm{makespan}}$ (ms) under different token lengths and compression settings.}
\label{tab:main_with_vectors}

\setlength{\tabcolsep}{2.6pt}
\renewcommand{\arraystretch}{1.12}
\scriptsize

\begin{tabular}{@{}>{\raggedright\arraybackslash}p{0.22\linewidth} ccc ccc ccc ccc ccc@{}}
\toprule
\multicolumn{1}{c}{\multirow{2}{*}{Slice token sizes}}
& \multicolumn{3}{c}{\shortstack{Compression factors\\$[0.5, 0.5, 0.50,$\\$ 0.50, 0.50, 0.5]$}}
& \multicolumn{3}{c}{\shortstack{Compression factors\\$[0.25, 0.25, 0.50,$\\$ 0.50, 0.50, 0.5]$}}
& \multicolumn{3}{c}{\shortstack{Compression factors \\$[0.25, 0.25, 0.80,$\\$ 0.60, 0.60, 0.6]$}}
& \multicolumn{3}{c}{\shortstack{Compression factors \\$[0.25, 0.25, 0.50,$\\$ 0.50, 0.80, 0.8]$}}
& \multicolumn{3}{c}{\shortstack{Compression factors \\$[0.5, 0.5, 0.50,$\\$ 0.50, 0.80, 0.8]$}} \\
\cmidrule(lr){2-4}\cmidrule(lr){5-7}\cmidrule(lr){8-10}\cmidrule(lr){11-13}\cmidrule(lr){14-16}
& \textbf{RTFS} & \textbf{PACS} & \textbf{Gain} & \textbf{RTFS} & \textbf{PACS} & \textbf{Gain} & \textbf{RTFS} & \textbf{PACS} & \textbf{Gain} & \textbf{RTFS} & \textbf{PACS} & \textbf{Gain} & \textbf{RTFS} & \textbf{PACS} & \textbf{Gain} \\
\midrule
$[128, 128, 128, 128, 128, 128] $   & 113.92 & 113.92 & 0.00\% & 105.92 & 104.92& +0.94\% & 114.92 & 116.92  & -1.74\% & 126.92  & 126.94 & -0.02\% & 134.92 & 132.43 & +1.85\% \\
$[64, 64, 128, 128, 256, 256]$   & 154.87 & 143.83 & +7.13\%  & 150.87 & 141.47 & +6.23\% &168.87 & 161.41 & +4.42\%  &191.87  & 181.47 & +5.42\%  & 196.87 & 184.59 & +6.24\% \\
$[64, 128, 256, 128, 192, 128]$   & 139.15 & 126.34 & +9.21\% & 132.15  & 132.15 & 0.00\%  & 145.15 & 136.53 & +5.94\%  & 157.15 & 148.54 & +5.48\%  & 168.15 & 155.52 & +7.51\% \\
$[512, 128, 256, 128, 192, 128]$   & 205.60 & 166.39 & +19.08\%  & 189.60 & 153.39 & +19.10\%  & 204.60 & 168.39 & +17.70\%  & 217.60 & 180.39 & +17.10\%  & 231.60 & 191.39 & +17.36\% \\
$[512, 128, 256, 384, 192, 64]$  & 230.33 & 193.66 & +15.92\%  & 213.33 & 179.66 & +15.78\%  & 230.33 & 192.66 & +16.36\%  & 236.33 & 199.66 & +15.52\%  & 236.33 & 199.66 & +15.52\% \\
\bottomrule
\end{tabular}
\end{table*}

\section{Conclusions}\label{sec:conclusions}
This paper investigated multimodal DNN workload orchestration in WNP to minimize the makespan of a unified communication–computation pipeline. To enable joint optimization of wireless transmission and DNN execution, we proposed O-WiN, a modular and scalable framework, and developed two heuristics, RTFS and PACS. Simulations revealed that PACS, which exploits pipeline parallelism between communication and computation, significantly reduces the end-to-end makespan compared to the sequential RTFS heuristic. Furthermore, PACS delivers larger gains under high modality heterogeneity, as early release and execution of modality-specific branches better hide wireless tail latency.

\ifCLASSOPTIONcaptionsoff
  \newpage
\fi

\end{document}